# Magnonic Combinatorial Memory for High-density Data Storage

Mykhaylo Balinskiy, Paulo Julio, Jeffrey Vargas, Diana Balaguer, and Alexander Khitun

*Department of Electrical and Computer Engineering, University of California - Riverside, Riverside, California, USA, 92521*

Correspondence to akhitun@engr.ucr.edu

**Abstract:** There is an urgent need to enhance the storage density of memory devices to accommodate the exponentially increasing amount of data generated by humankind. In this work, we describe Magnonic Combinatorial Memory (MCM), where the bits of information are stored in the signal propagation paths in the network. The number of paths among the elements of the network is much larger compared to the number of elements, which makes it possible to enhance the data storage density compared to conventional memory devices. MCM is an active ring circuit consisting of electric and magnonic parts. The electric part includes a broadband amplifier, phase shifters, and frequency filters. The magnonic part is a mesh of frequency-dependent elements. Signal propagation path(s) in the mesh depend on the amplitude/phase matching between the electric and magnetic parts. The operation of the MCM is described based on the network model, where information is encoded in the S-parameters of the network elements as well as in the element arrangement in the network. We present experimental data for MCM with a four-terminal magnonic element. The element consists of a single-crystal yttrium iron garnet $Y_3Fe_2(FeO_4)_3$ (YIG) film and magnets on top of the film. There are four micro antennas aimed to convert electromagnetic waves into spin waves and vice versa. One of the antennas is used as an input port while the other three are the output ports. Experimental data show the prominent dependence of the element S-parameters on the magnet arrangement. The number of possible arrangements scales factorially with the number of magnets. There is a number of bits that can be encoded into one magnet arrangement. It is possible to encode more than nine bits of information having only four magnets. The results demonstrate a robust operation of MCM with an On/Off ratio for path detection exceeding 50 dB at room temperature. Physical limits and practical constraints of MCM are also discussed.

I.       **Introduction**

We are witnessing an unprecedented increase in the amount of data generated by humankind. The global data is expected to exceed 175 zettabytes (ZB) by 2025, according to the International Data Corporation [1]. It is projected to be an over 20 % increase in the global data volume for every five years [2]. This exponential growth will shortly lead to the grand challenge for conventional storage systems, that may become unsustainable due to their limited data capacity, infrastructure cost, and power consumption [3]. For instance, it may not be enough silicon for flash-memory production by the year 2040 [4]. It stimulates the quest for novel memory devices with enhanced data storage density (i.e., the number of bits stored per area/volume/kilogram). *The number of bits stored in conventional memory devices is defined by the number of memory cells.* One cell can be in either two states (e.g., low resistance or high resistance). For the past decades, the enhancement in the storage density has been achieved by the miniaturization of the size of the memory cell. This trend is currently driving the research in nanometer-sized memory elements such as DNA-based [5] or sequence-defined macromolecules [6]. However, this approach may give only a temporary solution. It would be of great practical importance to develop a novel type of memory where the number of bits stored scales superlinear with the number of memory elements.

In our preceding work [7], we presented the concept of Magnonic Combinatorial Memory (MCM) and presented experimental data for the first prototype. The main idea of the combinatorial memory is to code information in the signal propagation paths in the mesh. The number of paths in the mesh is much larger compared to the number of elements in the mesh. It may be possible to drastically increase the data storage density by assigning the bits of information to the paths between the elements rather than using individual elements. Here, we further evolve the idea of combinatorial memory. The rest of the work is organized as follows. In the next Section II, we describe the network model for combinatorial memory. The network consists of frequency-dependent elements. The information is stored in the element characteristics as well as in the element arrangement in the network. In Section III, we present experimental data for magnonic combinatorial memory, where the non-linear element of the mesh is a ferrite film with magnets attached. The information is encoded in the mutual arrangement of magnets. The Discussion and Conclusions are presented in Sections IV and V, respectively.

## II. Combinatorial memory: the principle of operation

To explain the essence of combinatorial memory, we start with an active ring circuit shown in Fig.1 (A). It consists of a broadband amplifier $G(V)$, a frequency-dependent element $S(f)$, a voltage-tunable band-pass filter $f(V)$, a voltage-tunable phase shifter $\Psi(V)$, and a power sensor. For simplicity, we assume no attenuation/phase shift in the connecting wires and the power sensor. There are two conditions for auto-oscillations to occur in the circuit [8]:

$$G(V) + abs[S(f)] \geq 0, \qquad (1.1)$$

$$\Psi(V) + arg\,[S(f)] = 2\pi k, \text{where } k = 1,2,3, \ldots \qquad (1.2)$$

where $G(V)$ is the amplifier gain in decibels, $abs[S(f)]$ is the signal attenuation of the element in decibels, $\Psi(V)$ is the voltage-tunable phase shift, $arg\,[S(f)]$ is the phase shift to the propagating signal provided by the element. The first equation (1.1) states the amplitude condition for auto-oscillations: the gain provided by the broadband amplifier should be sufficient to compensate for losses in the element. The second equation states the phase condition for auto-oscillations: the total phase shift for a signal circulating through the ring circuit should be a multiple of $2\pi$. In this case, signals come in phase every propagation round. *Only signals propagating on the resonant frequency(s) that satisfy conditions (1.1) and (1.2) are amplified in the active ring circuit.* It takes just a few rounds of signal circulation in the ring circuit till the amplitude of the auto -auto-oscillations reaches the maximum.

In Fig.1(A), it is shown an active ring circuit with one-input and one-output frequency dependent element $S(f)$. The ports are marked as 1 and 2, respectively. The element can be described by the S-matrix as follows:

$$S(f) \equiv \begin{bmatrix} S(f)_{11} & S(f)_{12} \\ S(f)_{21} & S(f)_{22} \end{bmatrix}, \qquad (2)$$

where S-parameters are complex numbers that define the ratio between the amplitudes/phases of the incident and transmitted/reflected waves. The signal circulates clockwise in the circuit shown in Fig.1(A). In this case, $S(f)_{21}$ should be included in Eqs.(1.1-1.2) to define the conditions for auto-oscillations. This circuit can be considered as a memory, where the memory address is a combination of the filter bandpass frequency $f$ and the phase shift $\Psi$ of the phase shifter, while the memory state is the presence or absence of the auto-oscillations. The presence of auto-oscillation (i.e., high power circulating in the circuit) can be recognized as memory state 1. The absence of auto-oscillation (i.e., low power circulating in the circuit) can be recognized as memory state 0. The recognition is accomplished by the power sensor. The logic state is 1 if the output power exceeds some reference power $P_0$ and logic state is 0 otherwise. The information is encoded in the transmission characteristics of the frequency-dependent element (i.e., $S(f)_{21}$ parameter). It can be engineered such a way, that some of the auto-oscillation conditions are satisfied for certain combinations of $f(V)$ and $\Psi(V)$, and are not satisfied for other combinations.

In Fig.1(B) there is shown an example of the $S(f)_{21}$, where the red curve depicts the total amplitude $(G + abs[S(f)_{21}])$ and the blue curve depicts the $arg[S(f)_{21}]/2\pi$. The parameters are shown as function of frequency $f/f_0$, where $f_0$ is normalization frequency. We chose $abs[S(f)_{21}]$ to have a deep around $3f_0$ and $arg[S(f)_{21}]$ to scale proportional to $f^2$. We intentionally chose these dependencies to illuminate the feasibility of using both the amplitude and the phase for information encoding. Taking $\Psi(V) = \pi$ for all frequencies, one can see the combinations of $f$ and $\Psi$ leading to the auto-oscillations. The amplitude condition (1.1) is satisfied in all regions where the red curve is above 0. The green dash line in Fig.1(B) depicts the phase shift $\pi$. The phase condition is satisfied when the blue curve (the phase shift of the element) intersects with the green dashed line. For instance, combination $\{f_0, \pi\}$ results in the auto-oscillation as both amplitude and phase conditions are satisfied (i.e., depicted by the purple circle in Fig.1(B)). The phase condition is not met for the combination $\{2f_0, \pi\}$. Neither amplitude or phase conditions are satisfied for combination $\{3f_0, \pi\}$. The combination $\{4f_0, \pi\}$ results in the auto-oscillation as both amplitude and phase conditions are satisfied. Table I shows the Truth Table for the circuit in Fig.1. It is an example of five bits encoded in the one-input one-output element. It should be noted that the table is constructed for $\Psi(V) =$ $\pi$. There is a variety of ways to build other truth tables for the different frequency-phase combinations.

Next, we consider a circuit with a four-port frequency-dependent element as shown in Fig.2(A). There are two input ports marked as 1 and 2, and two output ports 3 and 4. The input ports are equipped with switches. There are voltage-tunable band-pass filters $f_i(V)$, a voltage-tunable phase shifter $\Psi_i(V)$, and power sensors $P_i$ at each output port. The element is described by the S-matrix as follows:

$$S(f) \equiv \begin{bmatrix} S(f)_{11} & S(f)_{12} & S(f)_{31} & S(f)_{41} \\ S(f)_{21} & S(f)_{22} & S(f)_{32} & S(f)_{42} \\ S(f)_{31} & S(f)_{32} & S(f)_{33} & S(f)_{43} \\ S(f)_{41} & S(f)_{42} & S(f)_{34} & S(f)_{44} \end{bmatrix} \quad (3)$$

The signal circulates clockwise as in the previous example in Fig.1. The following S-parameters: $S(f)_{31}, S(f)_{41}, S(f)_{32}$ and $S(f)_{42}$ define the presence/absence of the auto-oscillations. There is an important difference in the circuit operation compared to the previous one-input one-output port case. The number of possible paths (i.e., the path that satisfies the auto-oscillation condition) increases exponentially with the number of ports. For instance, the may be resonant paths 1-3, 2-3, 1-4, 2-4, or any combination of the above. Also, the presence/absence of the auto-oscillations depends on the combination of the input switches. There are three possible combinations: 01, 10, and 11, where 1 corresponds to the switch in the position On and 0 corresponds to the position Off. Combination 00 is not considered as the passive and active parts of the active ring circuit are not connected.

We want to outline the importance of signal interference for information encoding in the case of multiple input/output ports. In Fig.2(B), there are shown the results of numerical modeling showing the phase of the signal at output 3 for all configurations of the input switches. The blue, red, and green curves in Fig.2(B) correspond to the cases 01, 01, and 11. Thus, the blue curve correspond to $S(f)_{32}$; the red curve correspond to $S(f)_{31}$; the green curve corresponds to the superposition of the two signals. For simplicity, we choose $abs[S(f)] = 1$. $S(f)_{32}$ is chosen to be the same as the one in Fig.1(B). $S(f)_{31}$ is chosen to have a phase shit linearly proportional to the frequency. It is assumed that the frequency filters are set to the different frequencies $f_1 \neq f_2$. The green dashed line depicts the phase shift of $\pi$. The phase condition is satisfied when the curves intersect with the dashed line. There are the same combinations as in Fig.1(B) for the case 01 (i.e., the blue curve) as we took the same characteristics for the S-parameters. Combinations $\{2f_0, \pi\}$ and $\{4f_0, \pi\}$ results in the auto-oscillation for switch combination 10. Combinations $\{3f_0, \pi\}$, $\{4f_0, \pi\}$ and $\{5f_0, \pi\}$ result in the auto-oscillation for switch combination 11. Table II shows the correlation between the switch combination, frequency, phase, and the auto-oscillation power for output #3. Another table can be constructed for output #4 based on the chosen $S(f)_{41}$ and $S(f)_{42}$. Also, there are frequency-phase combinations for the case $f_1 = f_2$ where all four S-parameters come to play. It should be noted that Table II represents only a part of the Truth table for the four-terminal element. The size of the table (i.e., the number of possible combinations) increases exponentially with the numbers of possible frequencies and the number of phases.

Finally, we consider a network consisting of multiple multi-port frequency-dependent elements. For simplicity, it is shown a $4 \times 4$ network in Fig.3. The elements are connected to each other horizontally, vertically, and diagonally. There are eight input/output ports for each element. In general, a multi-port element with $m$ ports is described by the S-matrix as follows:

$$S(f) \equiv \begin{pmatrix} S_{11} & \cdots & S_{1m} \\ \vdots & \ddots & \vdots \\ S_{m1} & \cdots & S_{mm} \end{pmatrix} \quad (4)$$

In the case of a network (e.g., as shown in Fig.3) *all the $m \times m$ S-parameters* define the presence/absence of the auto-oscillations. The amplitude/phase of the S-parameters, as well as the mutual arrangement of the elements in the mesh, define the presence/absence of the auto-oscillations in the circuit. A general view of the truth table for a network is shown in Table III. Memory state is the combination of input switches, bandpass frequencies of the filters, and the combination of phase shifters. The state is the $m$ digit number, where $m$ is the number of outputs. Logic 1 corresponds to the auto-oscillations in the circuit (i.e., the power of oscillations exceeds some reference value $P_0$. Logic 0 corresponds to the absence of the auto-oscillations. The table is constructed for $\Psi(V) = \pi$.

In order to estimate the data capacity of the combinatorial memory, we consider a network with $m$ input ports and $m$ output ports. There are switches at each input port. There are frequency filters, phase shifters, and power detectors at the output ports. The memory address is the combination of switches, frequencies, and phases. The output is an $m-$ digit number that corresponds to the power at the output ports (i.e., 0 or 1 at $m$ ports). The total number of addresses can be estimated as following:

$$number\ of\ input\ combinations = (2^m - 1) \cdot 2^l \cdot \frac{k!}{(k-m)!}, \qquad (5)$$

where $l$ is the number of distinct phases per output port, and $k$ is the number of distinct frequencies. The first term on the right in Eq.(5) corresponds to the number of switch combinations. At least one switch should be in the position On. The second term in Eq.(5) corresponds to the number of phase combinations $\Psi_i \in \{\Psi_0, \Psi_1, .. \Psi_l\}$. The last term in Eq.(5) corresponds to the number of combinations to set $k$ distinct frequencies over $m$ output ports. The number of possible combinations is skyrocketing with the increase of network size. For instance, taking $m = 4$, $l = 2$, and $k = 4$, one obtains 1440 combinations. The number of combinations exceeds 1T for $m = 40$. Assuming that all of the input combinations can be utilized for information encoding, the maximum number of bits stored is given by:

$$max\ number\ of\ bits\ stored = (2^m - 1) \cdot 2^l \cdot \frac{k!}{(k-m)!} \cdot m. \qquad (6)$$

The data storage density of MCM is the number of bits stored normalized to the area of the device. The number of bits is given by Eq.(6). The area of the device $A$ scales with the number of ports $m$: $A = m^2 \cdot a$, where $a$ is the area of the single element. The data storage density can be estimated as follows:

$$data\ storage\ density = (2^m - 1) \cdot 2^l \cdot \frac{k!}{(k-m)!} / (m \cdot a). \qquad (7)$$

The data storage density of combinatorial memory increases superlinear with the size of the network. Theoretically, there are no limits for increasing the data storage density by exploiting a larger number of ports, phase combinations, and different frequencies. Practically, there are limitations on the precision of phase measurements, the bandwidth of the filters, etc. that will limit the number of possible input combinations (to be discussed later in the text). To use all or most of the possible memory addresses, there should be a sufficient number of element arrangements in the network. The classical formula for the number of possible combination for arranging distinct elements over possible places is given by [9]:

$$number\ of\ element\ arrangements = \frac{z!}{(z-y)!} \qquad (8)$$

where $z$ is the number of distinct elements, $y$ is the number of possible places in the network. The number of possible arrangements increases factorially with the number of distinct elements, where one arrangement corresponds to the number of signal propagation paths in the network. The key advantage of the combinatorial memory over the existing memory devices is the larger data storage density that scales factorially with the size of the mesh in contrast to the linear scaling in conventional memory.

### III. Experimental data

Combinatorial memory (i.e., as shown in Fig.3) can be realized in a variety of ways. For example, it may be all-optical or all-electrical structure. The frequency-dependent element is the key ingredient of combinatorial memory. The question is *How to build the frequency-dependent elements with a minimum number of components?* One of the possible approaches is associated with the use of spin waves (magnons) [7]. There are several appealing properties of the magnonic approach. (i) The element structure is pretty simple comprising ferrite film and magnets. (ii) Spin wave propagates much slower compared to electromagnetic waves, that makes it possible to achieve prominent phase shifts (e.g., up to $\pi$) in a compact structure [7]. (iii) The use of miniature magnets for spin wave path control inherits all the advantages of magnetic memory (i.e., no-volatility, scalability). In this part, we present experimental data illustrating the operation of Magnonic Combinatorial Memory, where the frequency-dependent element is a ferromagnetic film with magnets attached. It is possible to engineer the S-parameters with minimum number of components/devices by simply re-arranging the magnets on top of the film.

The cross-sectional view of the MCM device is shown in Fig. 4(A). It consists from the bottom to the top of a permanent magnet, Printed Circuit Board (PCB) with four antennas, a YIG film, Gadolinium Gallium Garnett (GGG) substrate, and a plastic plate with 16 pits for magnets to be inserted. The permanent magnet is a pair of commercially available NdFeB magnets (model BY0Y04 by K&J Magnets, Inc.) with the dimensions of 2.0" × 2.0" × 0.25". It is aimed to create a constant bias magnetic field. The bias field is about 375 Oe and directed in-plane on the YIG film surface. There are four micro antennas fabricated on PCB. The characteristic size of the antenna is 6 mm in length and 0.15 mm in width. These antennas are used as the input/output ports for spin wave excitation/detection. The ferrite film is made of YIG grown by liquid epitaxy on a GGG substrate. The film is not patterned. The thickness of the film is 42 μm. The saturation magnetization is close to 1750 G, the dissipation parameter (i.e., the width of the ferromagnetic resonance) ΔH = 0.6 Oe measured at 3 GHz. The YIG layer is placed on top of the PCB with antennas. The thickness of the GGG substrate is reduced by polishing to 0.3 mm from the initial 0.5 mm. There is a plastic layer mechanically placed over the GGG layer. There are 16 pits in the plastic layer for micro magnets to be inserted. The pits are located within the center of the structure between the micro antennas.

The schematics of the experimental setup are shown in Fig.4 (B). It is an active ring circuit comprising electric and magnetic parts. The magnetic part is the YIG film. One of the micro antennas marked as #1 is used to excite spin waves in the film. There is a phase shifter (Phase 1 in Fig.4(B)) connected in series with the input antenna. The other three micro antennas marked as #2, #3, and #4 are used to detect the inductive voltage produced by the propagating spin waves in the film. There is a bandpass frequency filter, a phase shifter, and a directional coupler for each output port. The frequency filters are commercially available YIG-sphere based voltage-tunable bandpass filters produced by Micro Lambda Wireless, Inc, model MLFD-40540. The filters at output ports #2, #3, and #4 are set to the central frequencies $f_1 = 1.614$ GHz, $f_2 = 1.838$ GHz, and $f_3 = 1.720$ GHz, respectively. These frequencies are chosen based on the preliminary measurements of the S parameters to ensure maximum signal transmission. The frequency bandwidth of the filters is about 15 MHz. Experimental data on the filter characteristics can be found in the Supplementary Materials. The frequency filters differentiate the frequency of the output signals coming to the different output ports.

The phase shifters are commercially available shifters from ARRA, model 9418 A. These phase shifters allow us to separately control the phase conditions Eq.1.2 for different paths (i.e., 1-2, 1-3, 1-4). The accuracy of the phase control is about $1^0$ in the operational frequency range. The directional couplers are KRYTAR, model 1820. The directional couplers are used to take a portion of of the output signal at each output port 2, 3, 4 as well as at the port 1 - the common part of the ring. The outputs are combined by the splitter (SPLT 1-3, Sigatek SP11R2F 1527) and applied to the broadband amplifier (three amplifiers Mini-Circuits, model ZX60-83LN-S+ connected in series). The power at each output port and the total power circulating in the ring circuit is measured by the spectrum analyzer (SA) GW Instek GSP-827. The spectrum analyzer is not required for the memory device operation. It is included in the experimental setup to collect the complete set of information about the frequencies of the auto-oscillations in the circuit.

In Fig.5, there is shown a collection of experimental data (i.e., S-parameters) measured for the film *without magnets*. The measurements are accomplished by VNA (see schematics Fig.4(B). The graphs in Fig.5 (A-C) show the $abs[S(f)_{21}]$, $abs[S(f)_{31}]$, and $abs[S(f)_{41}]$ measured separately, when only one output port was connected to the splitter. These data show the damping of the signal in the film in the frequency range from 1.5 GHz to 2.0 GHz. The black and the red curves correspond to the input power of -24 dBm and 0 dBm, respectively. The input power -24 dBm which is much less than the threshold of the nonlinear effects and spin wave amplitude saturations. The saturation starts as the input power exceeds -10 dBm. The input power of 0 dBm corresponds to the saturation. The saturation is due to the parametric excitation of the pairs of spin waves at half the oscillation frequency. In all further experiments, the input power of 0 dBm is taken as it satisfies the amplitude condition 1.1.

Fig.5 (D) show the $abs[S(f)_{2+3+4,1}]$, where all three outputs are connected in parallel and combined by the splitter. The data in Fig.5 (E) show the $arg[S(f)_{2+3+4,1}]$, where all three

outputs are connected in parallel. The measurements are accomplished by VNA connected between the phase shifter PhSh1 and port 1 of the device as it is shown in Fig. 4(B). The collected data give an insight of choosing the operational frequencies. It is logical to choose them around the peaks in transmission (see Fig. 5(D)). Ferrite film itself (i.e., without magnets) possesses complicated characteristics that define the presence/absence of auto-oscillations. Our objective is to demonstrate the ability to engineer these characteristics by placing magnets on top of the film. We used four micro-magnets made of NdFeB of volumes 0.02 mm$^3$, 0.03 mm$^3$, 0.045 mm$^3$, and 0.06 mm$^3$, respectively. These magnets are placed in the four corner pits. There are $4! = 4 \times 3 \times 2 \times 1 = 24$ possible magnet arrangements. Hereafter, we numerate these combinations by a number from 1 to 24. The complete list of the numbered magnet arrangements can be found in the Supplementary Materials.

In Fig.6, there is shown the collection of experimental data obtained for all 24 selected magnet configurations. In Fig.6(A), there are shown the frequencies of the auto-oscillation measured at the output ports. The black, the red, and the blue markers correspond to the output ports #2, #3, and #4, respectively. The observed frequencies are within the band-pass range of the filters. In Fig. 6(B), there are shown the output power at the different output ports. The black, the red, and the blue markers correspond to the output ports #2, #3, and #4, respectively. As one can see from Fig.6(B), auto-oscillations are observed at some of the magnet configurations and not present for the others. All phase shifters are set to $0\pi$. The difference between the On and Off states exceeds 50 dB at room temperature.

The combination of the band pass frequency and the phase serves as the memory address. In order to illustrate this idea, we performed a set of experiments for a different set of bandpass frequencies and different states of the phase shifters. The filters at output ports #2, #3, and #4 are set to the central frequencies $f_1 = 1.397$ GHz, $f_2 = 1.751$ GHz, and $f_3 = 1.547$ GHz, respectively. In Fig.7(A), there are shown the frequencies of the auto-oscillations obtained for different magnet configurations. The measurements are accomplished for all phase shifters set to $0\pi$. In Fig. 7(B), there are shown the output power at the different output ports. The black, the red, and the blue markers correspond to the output ports #2, #3, and #4, respectively. In Fig.7(C), there are shown the frequencies of the auto-oscillations obtained for different magnet configurations for the phase shifters set to $\pi/2$. The output power at the different output ports is shown in Fig7(D). As in the previous case shown in Fig.6, there is a big difference in the output power. The On/Off ratio exceeds 50 dB in all cases. We consider memory state 1 if the output power is above -20 dBm. The memory state is 0 if the output power is less than -70 dBm. The correlation between the magnet configuration and the output power is summarized in Table IV. The first column in Table IV shows the magnet configuration. The second column shows the memory states (three bits) obtained for $f_1 = 1.614$ GHz, $f_2 = 1.838$ GHz, $f_3 = 1.720$ GHz, and $\Phi_1 = \Phi_2 = \Phi_3 = 0\pi$. In the third column in Table IV, there are shown the memory states for $f_1 = 1.397$ GHz, $f_2 = 1.751$ GHz, $f_3 = 1.547$ GHz, and $\Phi_1 = \Phi_2 = \Phi_3 = 0\pi$. In the fourth column, there are shown the memory states for $f_1 = 1.397$ GHz, $f_2 = 1.751$ GHz, $f_3 = 1.547$ GHz, and $\Phi_1 = \Phi_2 = \Phi_3 = \pi/2$. As one can see, there is always an arrangement of magnet to meet any of the three-digit combinations. It should be noted that Table IV shows only a very small fraction of possible magnet configurations. There is $16^{16}$ possible arrangements for 16 magnets in 16 places (i.e., with the possibility of placing the same magnets in different pits). The collected data for 24 combinations demonstrates an example of programing the Magnonic Combinatorial Memory by the magnet arrangement.

In the previous section, it was assumed that the presence/absence of the auto-oscillations is defined by the S-parameters of the frequency-dependent element. In order to validate this hypothesis, we measured S-parameters for different magnet configurations and check the results with the data presented in Figs.6 and 7. In Fig.8, there are shown six curves of different color corresponding to the $S(f)_{21}$ measured at the six selected magnet configurations. The amplitude and the phase of $S(f)_{21}$ are shown in Fig.8(A) and Fig.8(B), respectively. The data are collected in the frequency range from 1.2 GHz to 1.9 GHz. Fig.8(C) shows the variation of the phase of $S(f)_{21}$ in the frequency range from 1.39 GHz to 1.44 GHz. The data show the prominent effect of magnet arrangement on the S-parameter. Indeed, changing the arrangement of magnets one can control the resonance conditions (i.e., the amplitude and the phase conditions of the auto-oscillations) for the whole circuit.

The comparison of the results obtained with the closed active ring circuit shown in Figs.6 and 7 and results obtained with VNA for the open circuit is shown in Table V. The comparison is made for the six arbitrarily selected magnet configurations. The zeros and ones in the column correspond to the expected absence and presence of the auto-oscillations. The data in the middle row are the predictions made on the basis of the measurements with the VNA. The results in the bottom row are the actual results measured in the active ring circuit. The results are quite consistent. However, there are two cases (i.e., marked with red color) when the conditions for auto-oscillation are predicted by the VNA measurements but not observed in the active-ring circuit configuration. This discrepancy may be explained by the additional phase shift introduced by the amplifier, non-linear effects, etc.

In Fig.9, there are shown experimental data: the amplitude and the phase of $S(f)_{21}$ measured for magnet configuration #1 at the output ports # 2 Fig.9(A) and # 3 Fig.9(B). The black curve in the graphs corresponds to the amplitude while the red curve corresponds to the phase of the S-parameter. The amplitude condition for the auto-oscillation is met for $absS(f)_{21} \geq 0$. The external phase shifter is set to $\Phi_1 = 0\pi$. The phase condition is met for $argS(f)_{21} = 0\pi$ (i.e., the green line in the graphs). According to the data in Fig.9(A), the output power at port #2 is low. The amplitude and the phase conditions are met at different frequencies. As a result, there are no auto-oscillations at the bandpass frequency of filter 1. The logic state of the memory is 0. In Fig.9(B), there are shown data for port #3 obtained for the same magnet configuration. Both the amplitude and the phase conditions for auto-oscillations are met in the same frequency range (i.e., the frequency range of the bandpass filter at the output #3). In this case, the memory state is 1. The experimental data presented in Fig.9 give an insight into the combinatorial memory programming. The set of experimental data for other magnet configurations is presented in the Supplementary Materials.

## IV.     Discussion

There are several observations we can make based on the obtained experimental data. (i) There is a prominent dependence between the arrangement of magnets on top of the ferrite film, frequency-phase combination and the presence/absence of the auto-oscillation. The rows in Table IV show such correlations experimentally verified.  (ii) The On/Off ratio (i.e., oscillations/no oscillations) exceeds 50 dB at room temperature (see Fig.4). The high On/Off ratio is a must for building robust memory devices.  The achieved high On/Off is due to the use of amplification (i.e., the broadband amplifier). It is interesting to note that the amplification in the active ring circuit is both amplitude and phase depended. It may provide an additional tool for building noise-immune memory devices. (iii)  One may conclude on the feasibility of engineering frequency-dependent elements for combinatorial memory by utilizing ferrite films with magnets attached. The structure of the element (i.e., ferrite film + magnet(s)) is pretty simple. At the same time, it gives us several degrees of freedom (e.g., placing magnets in different locations of the film, engineering magnet shape/size, placing a combination of magnets) to control the S-parameters. For instance, the data show the change in the amplitude and the phase shift of the propagating spin waves (see Fig.7) depending on the magnet arrangements. In turn, the modification of the S-parameters is quite large to control the auto-oscillation conditions (see Fig.8). Overall, the obtained results are in favor of magnonic combinatorial memory approach that may provide a fundamental increase in the data storage density compared to conventional memory devices.

There are some appealing properties of active ring circuit to be exploited in combinatorial memory. (i) It makes it possible to differentiate the signal propagation paths by the amplitude change and the accumulated phase. The more paths we can differentiate the more bits of information can be encoded. (ii) It is also possible to apply parallel path search in the active ring circuit. An active ring circuit starts with a superposition of signals on all frequencies propagating through all possible paths. Only the signals that met both the amplitude and phase conditions are amplified. A more detailed explanation can be found in Ref. [10].   (iii) The On/Off ratio may be further increased by optimizing the device structure (e.g., using elements with steeper dispersion characteristics). (iv) It takes less than one millisecond for the demonstrated device to reach saturation. The estimated power consumption is about 1 µJ per 3-bit retrieval in the demonstrated device. It can be further minimized by reducing the spin wave propagation distance. (v) There are some modifications in the device structure compared to the schematics of MCM shown in Fig.3. as well as from the original MCM structure as presented in [7]. The device is built on the base of unpatterned ferrite film. There are no sensors within the film to trace the spin wave routes. On the one hand, it reduces the number of bits to be read-out at a time. On the other hand, it makes the whole structure much simpler in fabrication in engineering.

The data encoding in combinatorial memory exploits both the amplitude and the phase characteristics of the element. The example of a one-input one-output element in Fig.1. shows that just a single element may store a number of bits. The data capacity increases with the non-linearity of the dispersion characteristics. The number of bits that can be stored increases linearly with the number of frequencies that can be used for independent data encoding. Wave interference is another degree of freedom that comes to play in multi-terminal elements. The elements can be engineered (i.e., the S-parameters) to have the desired response (i.e., the auto-oscillations) for different combinations of input switches. The number of switch combinations scales exponentially with the number of input ports. Finally, the assembling of multi-port elements in a network, as shown in Fig.3 offers an intriguing possibility to code information in the mutual arrangement of elements in a network. The number of possible arrangements scales factorially with the size of the network. Altogether it opens a big room for data storage density enhancement. In this work, only a small portion of all possible input combinations has been experimentally studied. For instance, we used only two sets of filter bandpass frequencies and two sets of phases. The increase in the number of frequency-phase combinations will exponentially increase the number of memory addresses to be used for information encoding. Using just a small fraction of these addresses will provide a fundamental advantage in data storage density.

The key question is related to the data storage density that can be practically achieved in MCM. There are three ways to increase the number of bits stored in MCM: increase the number of input/output ports, utilize more states for the phase shifters, and make use of a larger number of frequencies. The number of input combinations according to Eq.5 is skyrocketing with increasing of $m$, $l$, and $k$. There should be enough magnet arrangements to ensure independent information encoding for all combinations. The number of possible magnet arrangements according to Eq.6 scales factorially with the number of different elements z and the size of the network $m^2$. It promises a fundamental increase in the data storage density compared to conventional devices. At the same time, there are physical limits (e.g., the accuracy of measurements) that will limit the number of phases/frequencies that can be recognized. It should be noted that the experimental data collected in this work covers only a small percentage of possible combinations (e.g., magnet configurations, etc.) Is it possible to effectively control S-parameters by placing just one magnet in different locations? Would it be more efficient to use magnets of a special shape/size for S-parameter control (e.g., as studied in Ref.[11])? These and many other questions should be further addressed. There are several critical comments to mention. (i) The present work lacks the results of numerical modeling that may give insight into the practical limits of MCM. A high-fidelity model would require an enormous amount of work to link the properties of ferrite films and magnets to the S-parameters. (ii) Only a small part of the truth table has been experimentally verified. The data in Table IV show only results for the two sets of frequency filters and two phases. There are eight possible frequency combinations for each frequency set. It would take an enormous amount of time to check all possible frequency/phase combinations even for a relatively simple device demonstrated in this work. (iii) There are certain disadvantages of the proposed MCM compared to conventional RAM. The bits in the conventional RAM are separate independently addressable discrete bits. It takes one operation to change one bit. In contrast, it may take the re-arrangement of all magnets in MCM just to change one bit in the stored massive array of data. That is the price to pay for using a collective state for information storage. Read-only memory is the most promising application for the proposed MCM as the arrangement of magnets is fixed and is not modified after fabrication.

There are many practical applications requiring high-density ROMs [12]. This work aims to evolve the concept of MCM and outline its potential advantages.

## V. Conclusions

We considered magnonic combinatorial memory for high-density data storage. Its principle of operation is explained by a network model. A frequency-dependent element is the key ingredient of combinatorial memory. Some amount of information is encoded in the element attenuation and dispersion characteristics. The most of information is encoded in the element arrangement in the network. Signals propagating on different paths in the network can accumulate different phase shifts and attenuation. The information is extracted using the unique property of the active ring circuit to amplify signals that met amplitude and phase conditions. We presented experimental data showing the feasibility of building a frequency-dependent element on the basis of a ferrite film with magnets attached. The S-parameters of the element can be controlled by the arrangement of magnets. The results are obtained for the single-crystal yttrium iron garnet $Y_3Fe_2(FeO_4)_3$ (YIG) film with four magnets attached. We used only four out of sixteen pits for magnet location to show how the arrangement of magnets affect the S-parameters of the element. Some of the arrangements result in the auto-oscillations and other do not. The results demonstrate a robust operation with On/Off ratio exceeding 50 dB at room temperature. It may be possible to store more than 1kB of memory using just four magnets compared to four bits in conventional memory. The material structure and principle of operation of MCM are much more complicated compared to conventional memory. At the same time, MCM may pave the road to unprecedented data storage capacity. There are many questions regarding the most efficient ways of MCM engineering to be further explored.

**Methods**

**Device Fabrication**

The core of the device is made of single-crystal $Y_3Fe_2(FeO_4)_3$ film. The film was grown on top of a (111) Gadolinium Gallium Garnett ($Gd_3Ga_5O_{12}$) substrate using the liquid-phase epitaxy technique. The thickness of the film is 42 μm. The saturation magnetization is close to 1750 G, the dissipation parameter (i.e., the width of the ferromagnetic resonance) ΔH = 0.6 Oe. The bias magnetic field is provided by the permanent magnet made of NdFeB.

**Measurements**

The excitation and detection of spin waves in the ferrite film were accomplished by four short-circuited antennas. The antennas are connected to a programmable network analyzer (PNA) Keysight N5241A. The filtering is by the commercially available filters produced by Micro Lambda Wireless, Inc, model MLFD-40540.

**Data availability**

All data generated or analyzed during this study are included in this published article and the Supplementary Materials.

**Author contributions**

M.B. built the prototype and accomplished experiments. P.J. assisted with data acquisition. J.V. and D.B. assisted with data arrangement. A.K. conceived the idea of combinatorial memory. All authors wrote and reviewed the manuscript.

**Competing financial interests**

The authors have no financial or non-financial conflicts of interests.

**Acknowledgments**

This work was supported by the National Science Foundation under grant # 2423929 and is supported in part by funds from federal agency and industry partners as specified in the Future of Semiconductors (FuSe) program.

**Figure Legends**

**Figure 1:** (A) Schematics of the active ring circuit shown that consists from a broadband amplifier $G(V)$, a two-port frequency-dependent element $S(f)$, a voltage-tunable band pass filter $f(V)$, a voltage-tunable phase shifter $\Psi(V)$. (B) An example of the $S(f)_{21}$, where the red curve depicts the $abs[S(f)_{21}]/G_0$ ($G_0$ is the amplification provided by the amplifier) and the blue curve depicts the $arg[S(f)_{21}]/2\pi$. The parameters are shown as function of frequency $f/f_0$, where $f_0$ is normalization frequency. The blue circles depict the points in the frequency-phase plot where the phase conditions of auto-oscillations are satisfied. The red circle around the blue circle means that both the amplitude and the phase conditions for auto-oscillations are satisfied.

**Table I:** Truth Table for the circuit in Fig.1. Logic 1 corresponds to the auto-oscillations in the circuit (i.e., the power of oscillations exceeds some reference value $P_0$. Logic 0 corresponds to the absence of the auto-oscillations. The table is constructed for $\Psi(V) = \pi$.

**Figure 2:** (A) Schematics of the active ring circuit with four-terminal frequency-dependent element $S(f)$. There are two input ports marked as 1 and 2, and two output ports 3 and 4. The input ports are equipped with switches.

**Table II:** Truth Table for the circuit in Fig.2. Logic 1 corresponds to the auto-oscillations in the circuit (i.e., the power of oscillations exceeds some reference value $P_0$. Logic 0 corresponds to the absence of the auto-oscillations. The table is constructed for $\Psi(V) = \pi$.

**Figure 3:** Schematics of a $4 \times 4$ network consisting of multi-port frequency-dependent elements. There are eight input/output ports for each element. The elements are connected to each other horizontally, vertically, and diagonally.

**Table III:** Truth Table for the circuit in Fig.3. Memory state is the combination of input switches, bandpass frequencies of the filters, and the combination of phase shifters. The state is the $n$ digit number, where $n$ is the number of outputs. Logic 1 corresponds to the auto-oscillations in the circuit (i.e., the power of oscillations exceeds some reference value $P_0$. Logic 0 corresponds to the absence of the auto-oscillations. The table is constructed for $\Psi(V) = \pi$.

**Figure 4:** (A) The cross-sectional view of MCM prototype. It consists from the bottom to the top of a permanent magnet, Printed Circuit Board (PCB) with four antennas, a YIG film, Gadolinium Gallium Garnett (GGG) substrate, and a plastic plate with 16 pits for magnets to be inserted. The ferrite film is made of YIG grown by liquid epitaxy on a GGG substrate. There are four micro antennas fabricated on PCB. The characteristic size of the antenna is 6 mm in length and 0.15 mm in width. These antennas are used as the input/output ports for spin wave excitation/detection. The thickness of the YIG film is 42 μm. The saturation magnetization is close to 1750 G, the dissipation parameter (i.e., the half-width of the ferromagnetic resonance) ΔH = 0.6 Oe measured at 3 GHz. (B) The schematics of the experimental setup. It is an active ring circuit comprising electric and magnetic parts. The magnetic part is the YIG film with four micro-antennas. One of the micro antennas marked as #1 is to excite spin waves. The other three micro antennas marked #2, #3, and #4 are the output antennas that are used to detect the inductive voltage produced by the propagating spin waves in the film. There are frequency filters (marked as F), phase shifters (marked as PhSh), directional couplers (marked as DC). Also, there is a Vector Network Analyzer (marked as VNA), and a Spectrum Analyzer (marked as SA).

**Figure 5:** (A-C) $abs[S(f)_{21}]$, $abs[S(f)_{31}]$, and $abs[S(f)_{41}]$ measured separately, when only one output port was connected to VNA. The black and the red curves correspond to the input power of -24 dBm and 0 dBm, respectively. (D) $abs[S(f)_{2+3+4,1}]$, where all three outputs are connected in parallel. (E) $arg[S(f)_{2+3+4,1}]$, where all three outputs are connected in parallel.

**Figure 6:** Collection of experimental data obtained for 24 selected magnet configurations. (A) The frequencies of the auto-oscillation measured at the output ports. The black, the red, and the blue markers correspond to the output ports #2, #3, and #4 respectively. (B) Output power at the different output ports. The black, the red, and the blue markers correspond to the output ports #2, #3, and #4 respectively. The data are collected for the following bandpass filter frequencies: $f_1 = 1.614$ GHz, $f_2 = 1.838$ GHz, and $f_3 = 1.720$ GHz, respectively. All phase shifters are set to $0\pi$.

**Figure 7:** Collection of experimental data obtained for 24 selected magnet configurations for the different set of the filter bandpass frequencies: $f_1 = 1.397$ GHz, $f_2 = 1.751$ GHz, and $f_3 = 1.547$ GHz. (A) and (B) the frequency of the auto-oscillations and output power measured for phase shifters are set to $0\pi$. (C) and (D) the frequency of the auto-oscillations and output power measured for phase shifters are set to $0\pi/2$.

**Table IV:** Truth table obtained for the two sets of bandpass frequencies and two sets of the phase shifters. The first column shows the magnet configuration. The second column shows the memory states (three bits) obtained for the first frequency set ($f_1 = 1.614$ GHz, $f_2 = 1.838$ GHz, $f_3 = 1.720$ GHz), and $\Phi_1 = \Phi_2 = \Phi_3 = 0\pi$. The third column shows the memory states for the second frequency set ($f_1 = 1.397$ GHz, $f_2 = 1.751$ GHz, $f_3 = 1.547$ GHz), and $\Phi_1 = \Phi_2 = \Phi_3 = 0\pi$. The fourth column shows the memory states for the second set of frequencies ($f_1 = 1.397$ GHz, $f_2 = 1.751$ GHz, $f_3 = 1.547$ GHz), and $\Phi_1 = \Phi_2 = \Phi_3 = \pi/2$.

**Figure 8:** Experimental data showing the modification of the S-parameters for different magnet configurations. The six curves of different color corresponding to the $S(f)_{21}$ measured at the six selected magnet configurations. (A) The amplitude of $S(f)_{21}$. (B) The phase of $S(f)_{21}$. (C) The enlarged view of $S(f)_{21}$ variation for different magnet configurations in the frequency range from 1.39 GHz to 1.44 GHz.

**Table V:** Comparison between the device output predicted by passive measurements by VNA and the actual experimental data in the active ring circuit. The data in the middle row are the predictions made on the basis of the measurements with the VNA. The results in the bottom row are the actual results measured in the active ring circuit.

**Figure 9:** (A) The amplitude and the phase of $S(f)_{21}$ for magnet configuration #1 at the output ports # 2. (B) The amplitude and the phase of $S(f)_{21}$ for magnet configuration #1 at the output ports # 3. The black curve in the graphs corresponds to the amplitude while the red curve corresponds to the phase of the S-parameter. The amplitude condition for the auto-oscillation is met for $abs S(f)_{21} \geq 0$. The external phase shifter is set to $\Phi_1 = 0\pi$. The phase condition is met for $arg S(f)_{21} = 0\pi$ (i.e., the green line in the graphs).

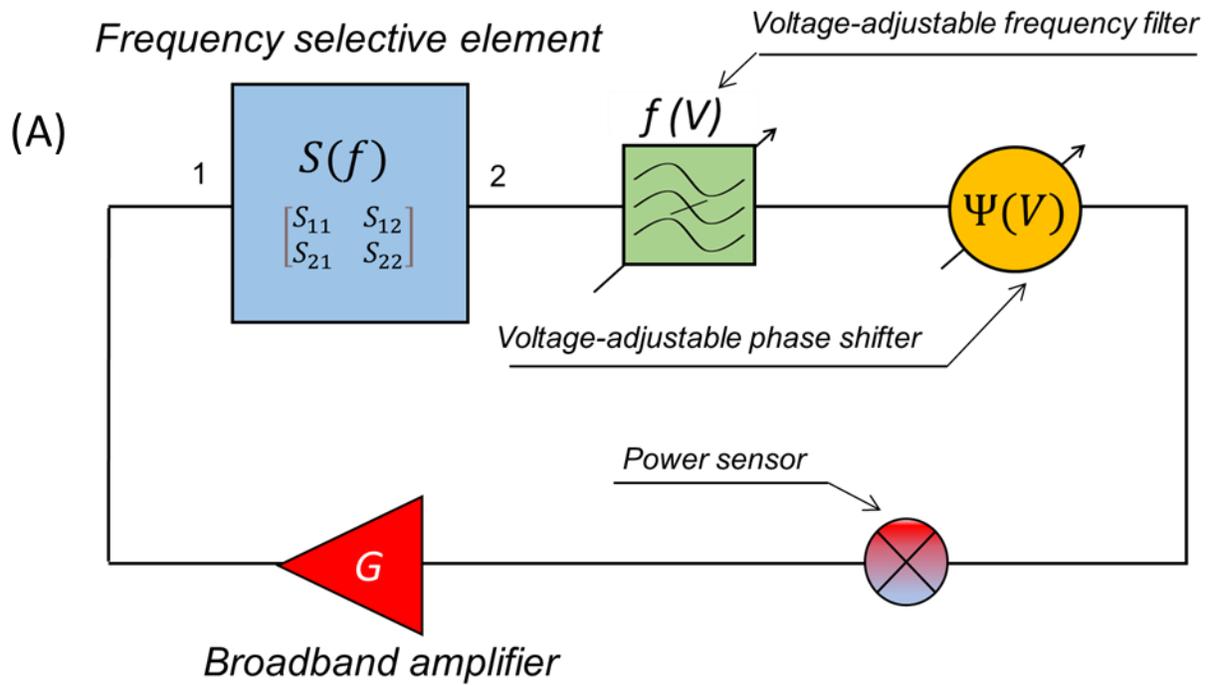

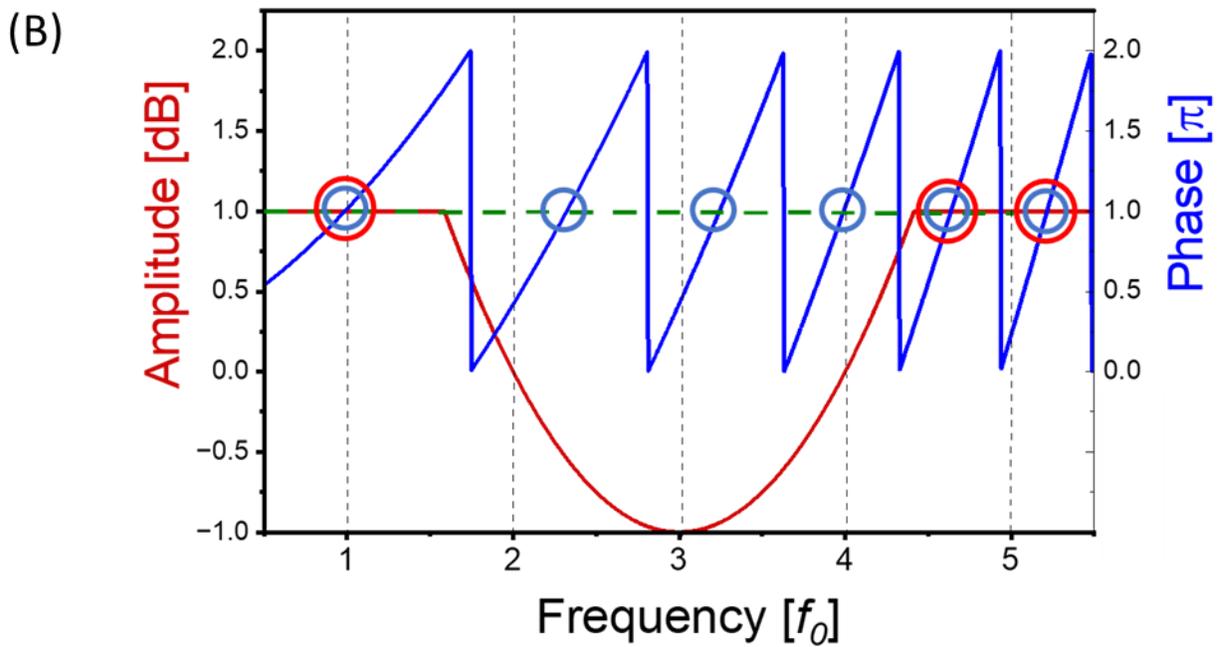

**Figure 1**

| $f$ | $\Psi$ | Auto-oscillation power |
|---|---|---|
| $f_0$ | $\pi$ | 1 |
| $2f_0$ | $\pi$ | 0 |
| $3f_0$ | $\pi$ | 0 |
| $4f_0$ | $\pi$ | 1 |
| $5f_0$ | $\pi$ | 0 |

**Table I**

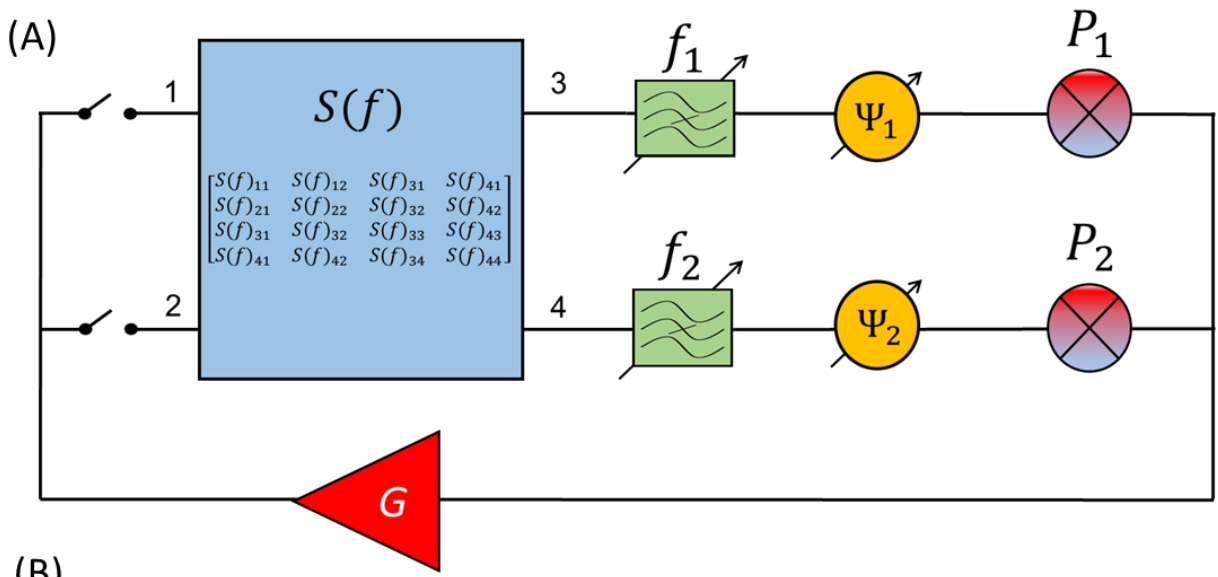
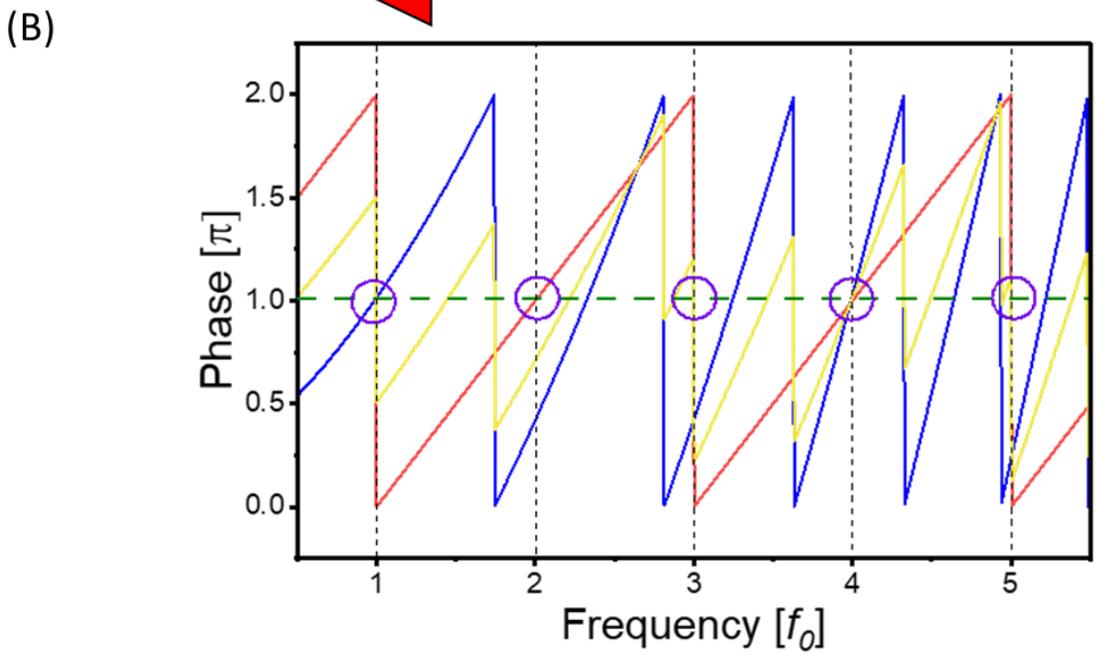

**Figure 2**

| Switch combination | $f$ at port 3 | $\Psi$ at port 3 | Auto-oscillation power |
|---|---|---|---|
| 01/10/11 | $f_0$ | $\pi$ | 1/0/0 |
| 01/10/11 | $2f_0$ | $\pi$ | 0/1/0 |
| 01/10/11 | $3f_0$ | $\pi$ | 0/0/1 |
| 01/10/11 | $4f_0$ | $\pi$ | 1/1/1 |
| 01/10/11 | $5f_0$ | $\pi$ | 0/0/1 |

**Table II**

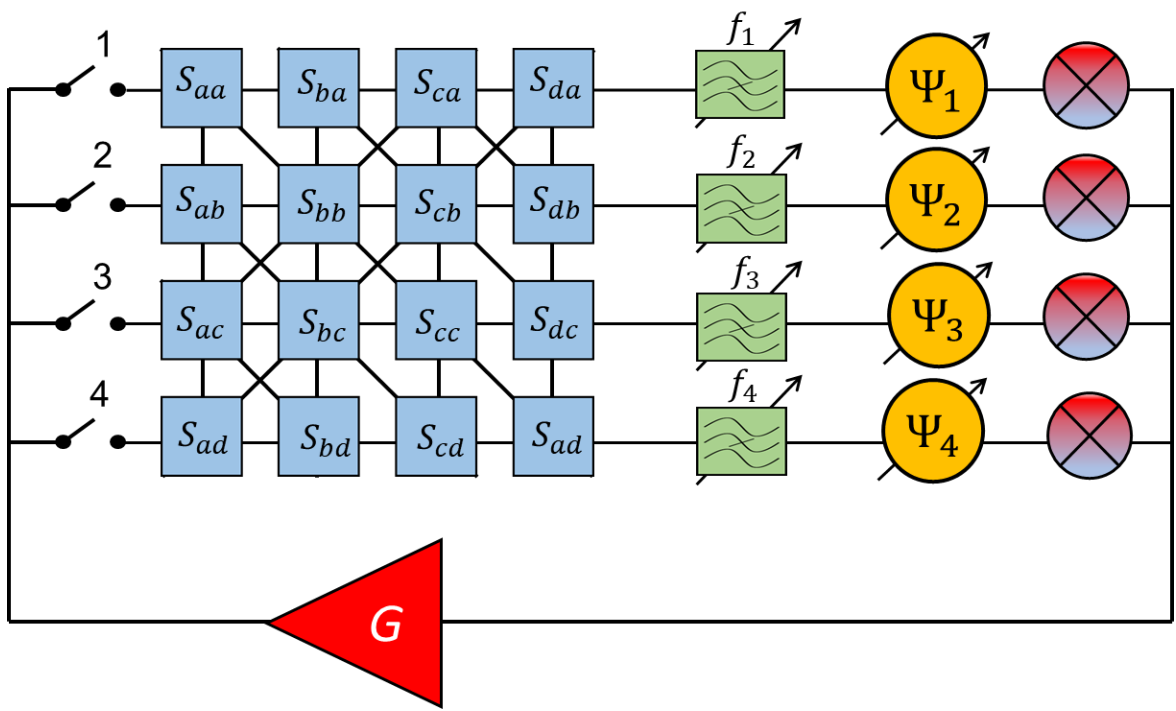

**Figure 3**

| Memory Address | | | Memory State |
|---|---|---|---|
| Switches | Frequencies | Phases | Power at the output $m$ ports |
| 000…1 | $\{f_1, f_2, ..f_k\}$ | $\{\Psi_1, \Psi_2, ..\Psi_l\}$ | 100..1 |
| ⋮ | ⋮ | ⋮ | ⋮ |
| 111…1 | $\{f_k, f_k, ..f_k\}$ | $\{\Psi_l, \Psi_l, ..\Psi_l\}$ | 011…1 |

**Table III**

(A)

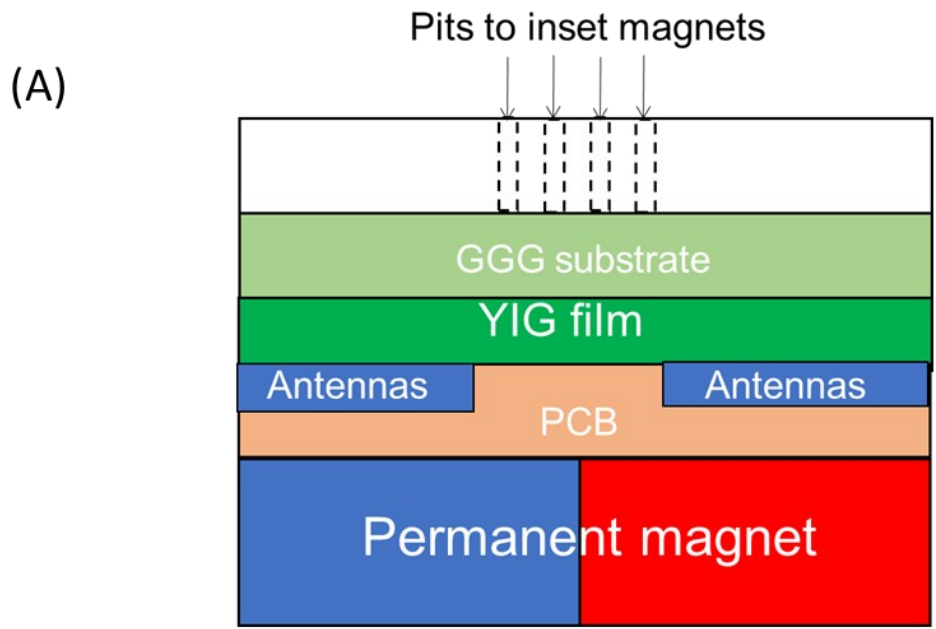

(B)

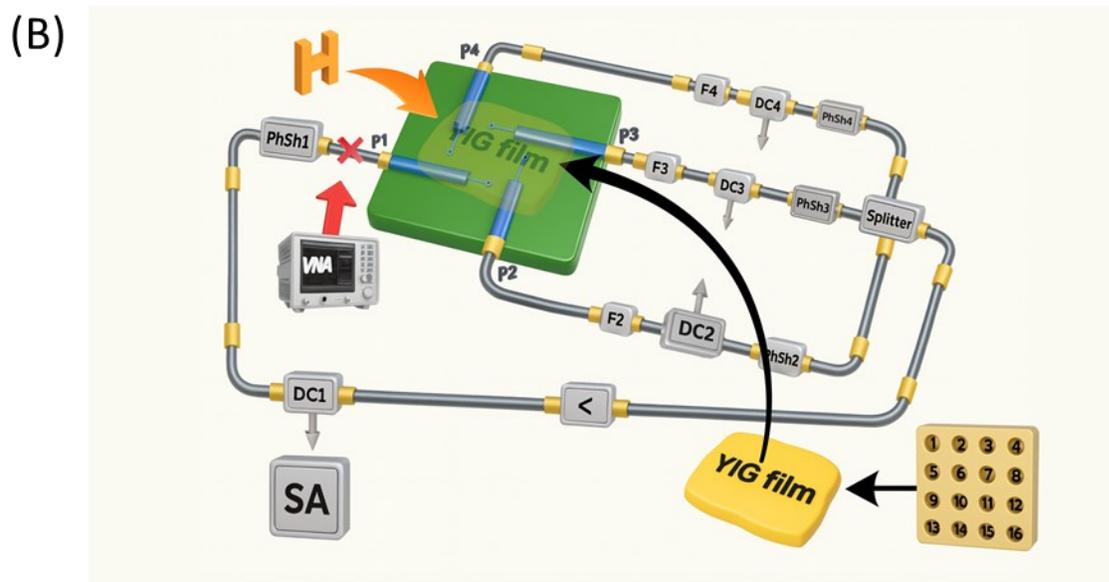

**Figure 4**

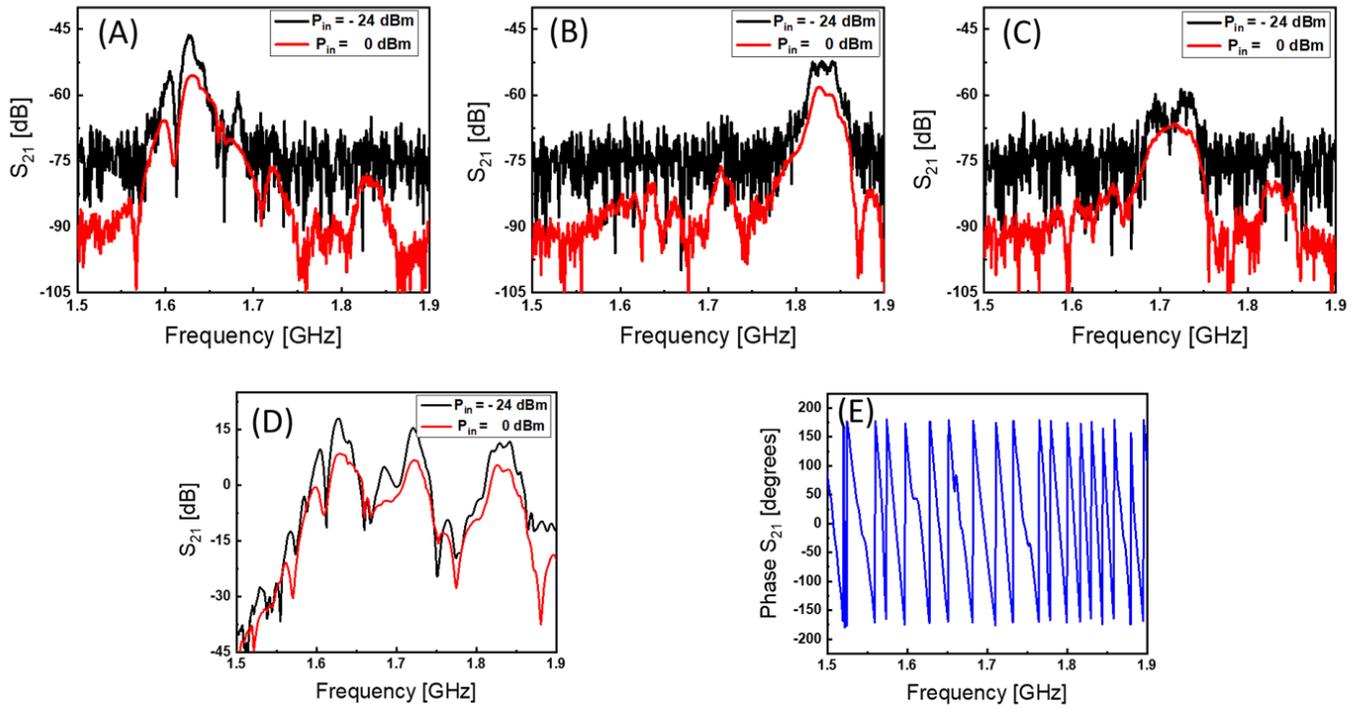

**Figure 5**

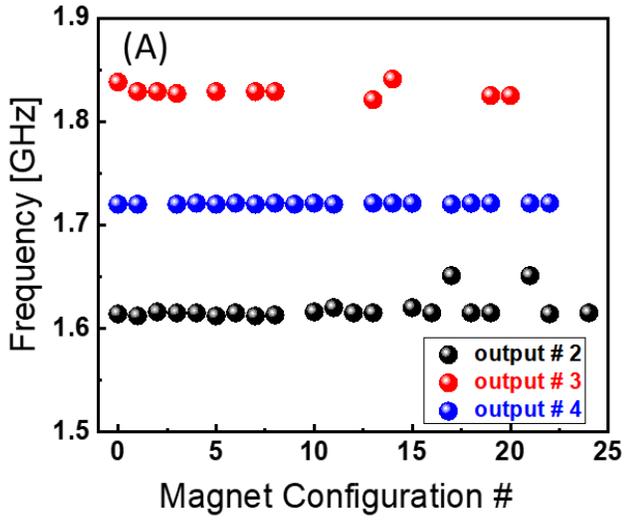 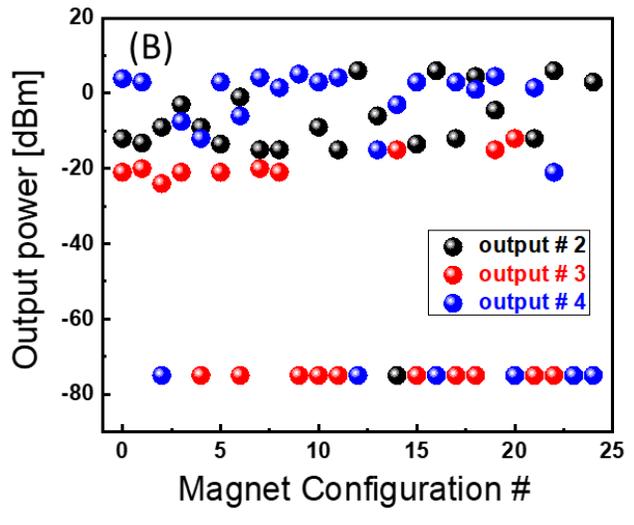

**Figure 6**

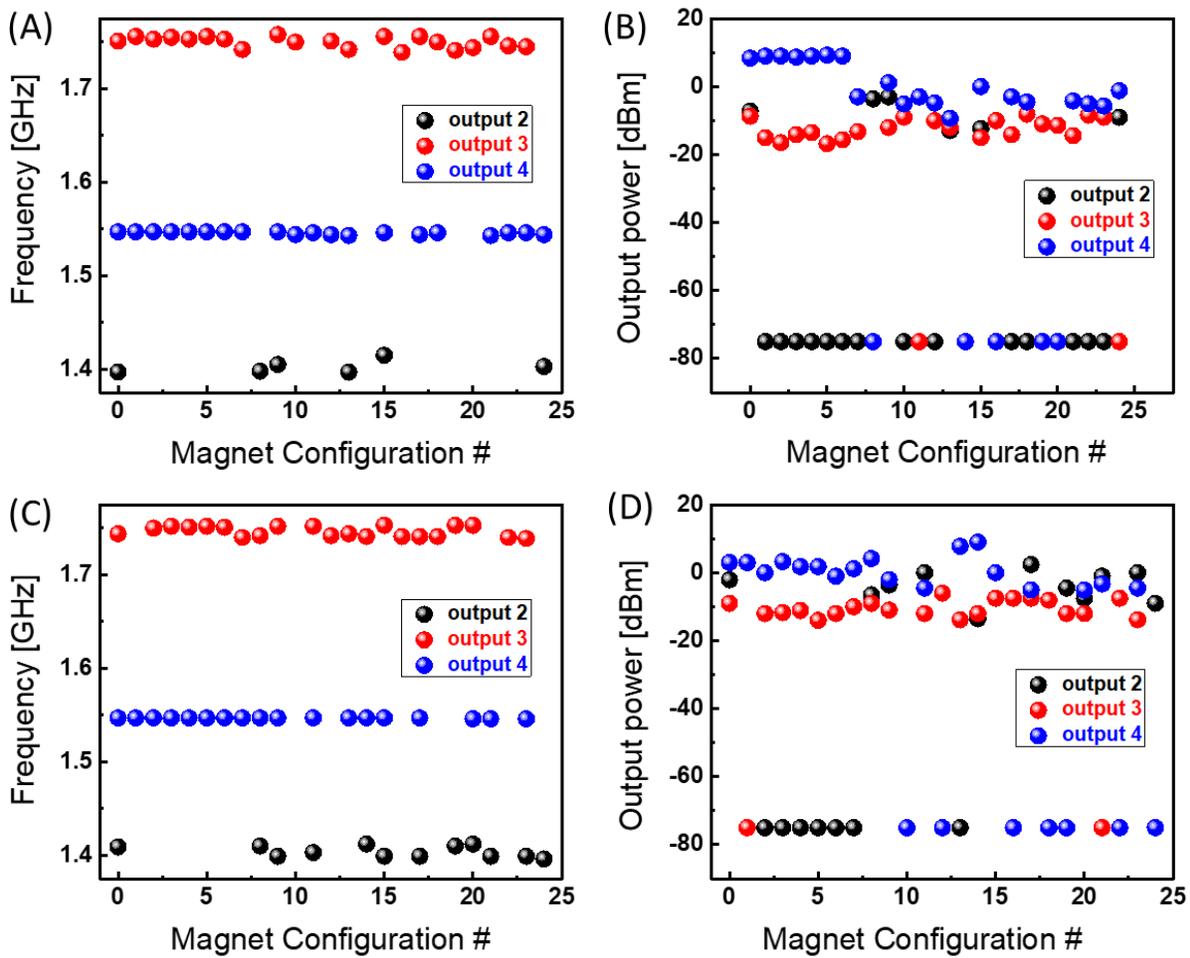

**Figure 7**

| Magnet Configuration # | Memory state Frequency set 1 $Phases = 0\pi$ | Memory state Frequency set 2 $Phases = 0\pi$ | Memory state Frequency set 2 $Phases = \pi/2$ |
|---|---|---|---|
| 1 | 111 | 011 | 001 |
| 2 | 110 | 011 | 011 |
| 3 | 111 | 011 | 011 |
| 4 | 101 | 011 | 011 |
| 5 | 111 | 011 | 011 |
| 6 | 101 | 011 | 011 |
| 7 | 111 | 011 | 011 |
| 8 | 111 | 100 | 111 |
| 9 | 001 | 111 | 111 |
| 10 | 101 | 011 | 000 |
| 11 | 101 | 001 | 111 |
| 12 | 100 | 011 | 010 |
| 13 | 111 | 110 | 011 |
| 14 | 011 | 000 | 111 |
| 15 | 101 | 111 | 111 |
| 16 | 100 | 010 | 010 |
| 17 | 101 | 011 | 111 |
| 18 | 101 | 011 | 010 |
| 19 | 111 | 010 | 110 |
| 20 | 010 | 010 | 111 |
| 21 | 101 | 011 | 101 |
| 22 | 101 | 011 | 010 |
| 23 | 000 | 011 | 111 |
| 24 | 100 | 101 | 100 |

**Table IV**

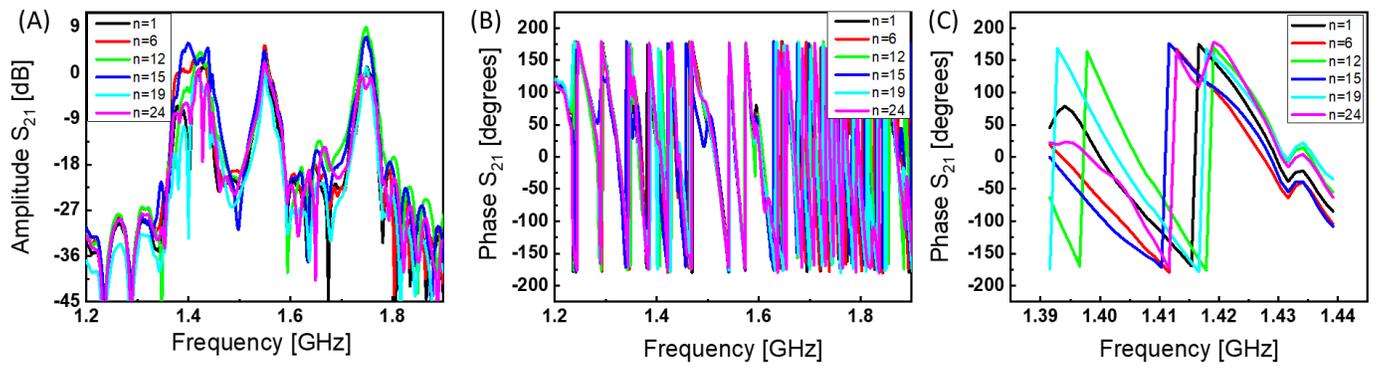

**Figure 8**

|  | # 1 | # 6 | # 12 | # 15 | # 19 | # 24 |
|---|---|---|---|---|---|---|
| Passive VNA | 011 | 011 | 111 | 111 | 010 | 001 |
| Active ring | 011 | 011 | 011 | 111 | 010 | 101 |

**Table V**

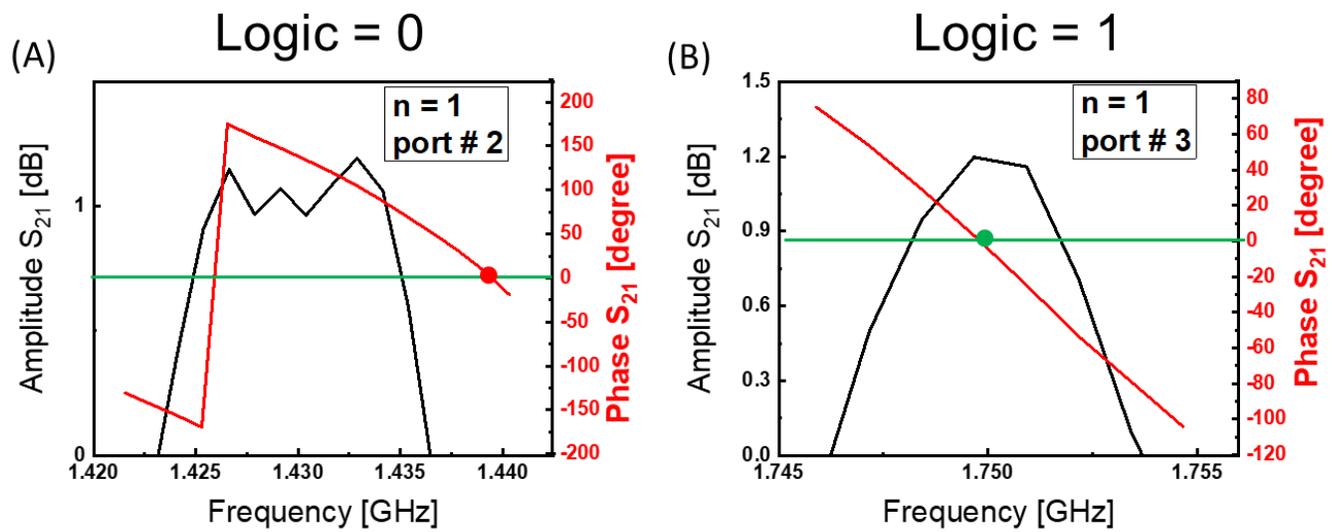

**Figure 9**